\documentclass[]{spie}  

\usepackage{amsmath,amsfonts,amssymb}
\usepackage{graphicx}
\usepackage[colorlinks=true, allcolors=blue]{hyperref}

\title{Real-time processing of the imaging data from the network of Las Cumbres Observatory Telescopes using BANZAI}

\author[a]{Curtis McCully}
\author[a]{Nikolaus H. Volgenau}
\author[a]{Daniel-Rolf Harbeck}
\author[a]{Tim A. Lister}
\author[a]{Eric S. Saunders}
\author[a]{Monica L. Turner}
\author[a]{Robert J. Siverd}
\author[a]{Mark Bowman}
\affil[a]{Las Cumbres Observatory, 6740 Cortona Drive, Suite 102, Goleta, CA 93117-5575, USA}
\authorinfo{Further author information: (Send correspondence to C.V.M.)\\C.V.M.: E-mail: cmccully@lco.global, Telephone: 1 805 880 1600}
\begin{document} 
\maketitle

\begin{abstract}
Work in time-domain astronomy necessitates robust, automated data processing pipelines that operate in real time. We present the BANZAI pipeline which processes the thousands of science images produced across the Las Cumbres Observatory Global Telescope (LCOGT) network of robotic telescopes each night. BANZAI is designed to perform near real-time preview and end-of-night final processing for four types of optical CCD imagers on the three LCOGT telescope classes. It performs instrumental signature removal (bad pixel masking, bias and dark removal, flat-field correction), astrometric fitting and source catalog extraction. We discuss the design considerations for BANZAI, including testing, performance, and extensibility. BANZAI is integrated into the observatory infrastructure and fulfills two critical functions: (1) real-time data processing that delivers data to users quickly and (2) derive metrics from those data products to monitor the health of the telescope network. In the era of time-domain astronomy, to get from these observations to scientific results, we must be able to automatically reduce data with minimal human interaction, but still have insight into the data stream for quality control.
\end{abstract}



\section{Introduction}
\label{sect:intro}  
In the era of time-domain astronomy, we need to rethink the way that we handle observational data. The traditional procedure of taking data manually, visually inspecting every image, and spending significant a amount of time and effort to produce final reduced products does not scale to upcoming surveys such as the Zwicky Transient Facility\cite{ZTF} (ZTF) and the Large Synoptic Survey Telescope\cite{LSST} (LSST), which will produce hundreds of thousands to millions of alerts per night.

The volume of data is not the only consideration. We are starting to find new classes of astrophysical sources that vary on a vast range of time scales: from seconds for fast radio bursts \cite{Katz2016}, to minutes for gamma-ray burst afterglows, to hours for kilonovae \cite{LIGO2017}, to days for supernovae. Thus, work in time-domain astronomy requires fast-turnaround data reduction without human intervention.

Las Cumbres Observatory Global Telescope\cite{Brown2013} (LCOGT) is pioneering the future of time domain follow-up. We currently operate a network of 21 small-aperture telescopes around the world which produce nearly 50,000 images ($\sim 1.5$ TB uncompressed raw data) per month. Rather than acting individually, our telescopes operate in unison. Specifically, users submit requests to a single access point, which are then dynamically scheduled across the network resources\cite{Saunders2014}. From a user's perspective, LCOGT is a single facility, so it is necessary to produce homogeneous data for heterogeneous instruments by removing the instrumental signatures across the whole network of telescopes. 

To meet these challenges we have produced the ``Beautiful Algorithms to Normalize Zillions of Astronomical Images'' (BANZAI) pipeline\cite{banzai}. BANZAI reduces all of the images taken with LCOGT in real time. Below, we present the design choices we made for the BANZAI pipeline and how the pipeline is integrated into the overall dataflow of the observatory. Finally, we discuss the lessons LCOGT has learned which illustrate the challenges of running a high volume data processing service.

\section{Data flow from telescope to user}
LCOGT provides three tiers of data to users: raw, preview, and processed. The ``raw'' files contain unprocessed, 16-bit image data exactly as it was downloaded from the CCD. The ``preview'' and ``processed'' files have been calibrated to remove the instrument signature and also include source catalogs as additional extensions. Although both preview and processed files follow the same procedure, they may differ in which calibration files were used. ``Preview'' files are produced on-the-fly using recent calibration data, often from the previous day. ``Processed'' files are produced at the end of each night, using master calibration files from the same day as the observations (if available). All of our files are provided in Rice-compressed, ``fpacked''\cite{fpack1,fpack2} format.

The interaction between users and the LCOGT telescopes is significantly different than with traditional facilities. For classical observing, an astronomer would travel to the telescope, observe through the night, and take their data home on an external hard drive (or even DAT tapes) to reduce. Processing data from a single observational run often required significant time and effort. Queue observing solves the issue of needing to travel to the telescope which makes rapid follow-up possible. However, many queue scheduled observatories still require manual data reduction. Both rapid response and automatic, robust data reduction are necessary to follow up alerts from upcoming surveys like ZTF and LSST. 

\begin{figure}
\begin{center}
\includegraphics[width=\textwidth]{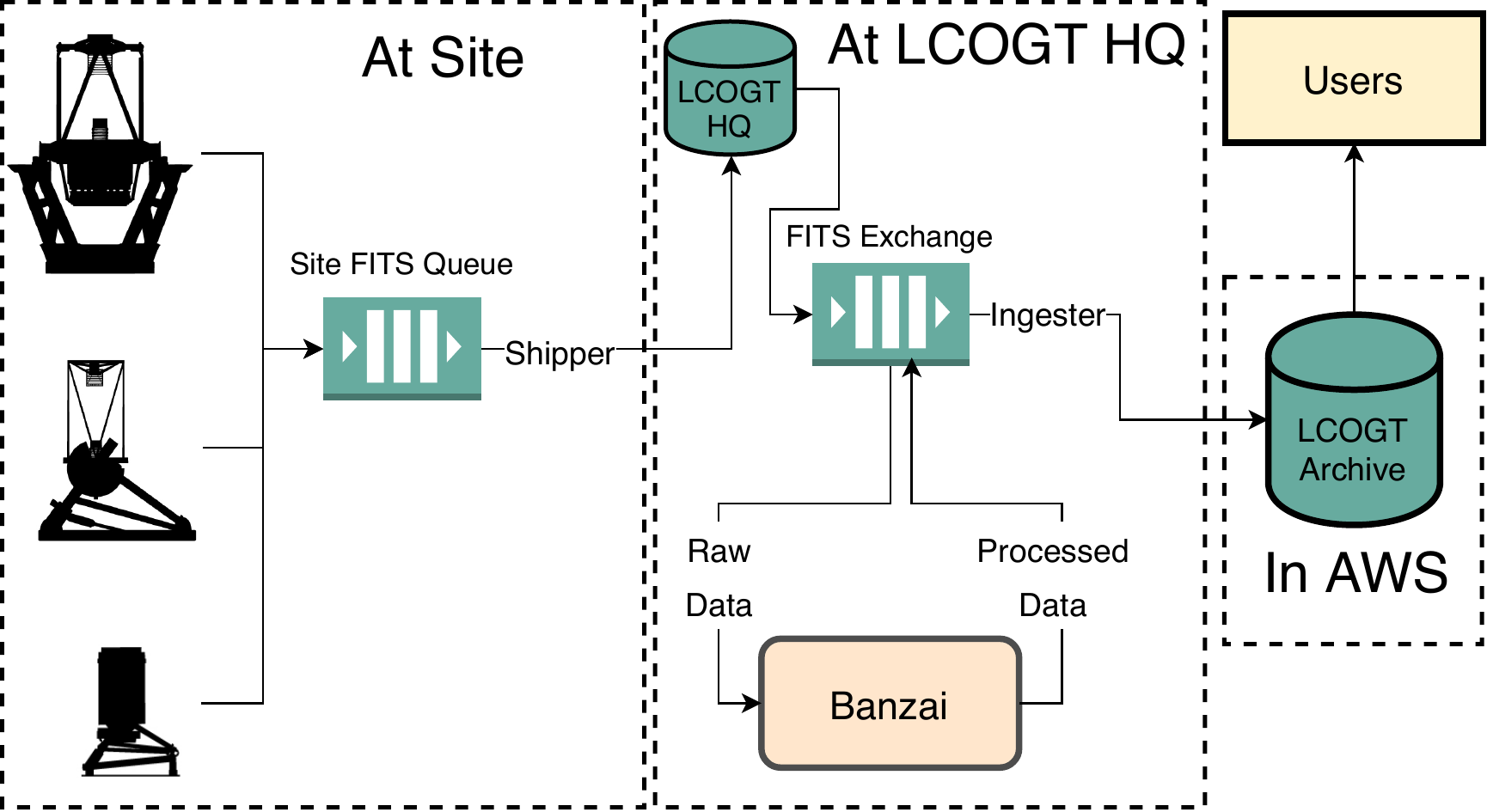}
\end{center}
\caption 
{ \label{fig:dataflow} Schematic of the flow of data from the end of exposure to being available to users. Once the shutter closes, the FITS file is placed on the queue to be shipped back to the LCOGT headquarters. After it arrives, BANZAI pulls it off the FITS Exchange and processes the data. After the data is processed it gets ingested into the archive (hosted in the cloud) where it can be downloaded by users.} 
\end{figure} 

Instead, LCOGT dynamically responds to changes in the system. When a user submits a request, a dynamic scheduler then optimizes when the observation should be attempted and at which site\cite{Saunders2014}. 

Figure \ref{fig:dataflow} illustrates the dataflow after an observation is taken. Once the shutter closes, the telescope writes a FITS file that is ``fpacked''\cite{fpack1,fpack2} to on-site storage. The name and location of this file are added to a transfer queue (we use RabbitMQ\cite{rabbitmq}, but we are not limited to this choice of implementation). The ``Shipper'' then transfers the file back to the LCOGT headquarters, and puts it on another queue, the ``FITS Exchange'' (again implemented using RabbitMQ\cite{rabbitmq}). The raw frames are then pushed to the science archive, which is hosted in the cloud using Amazon Web Services\cite{aws}, where they can be downloaded by users. BANZAI also listens to the FITS Exchange, reducing images as they are added running in a ``preview'' mode. These processed images are then placed back on the FITS Exchange to be ingested into the science archive. 

\begin{figure}
\begin{center}
\includegraphics[width=\textwidth]{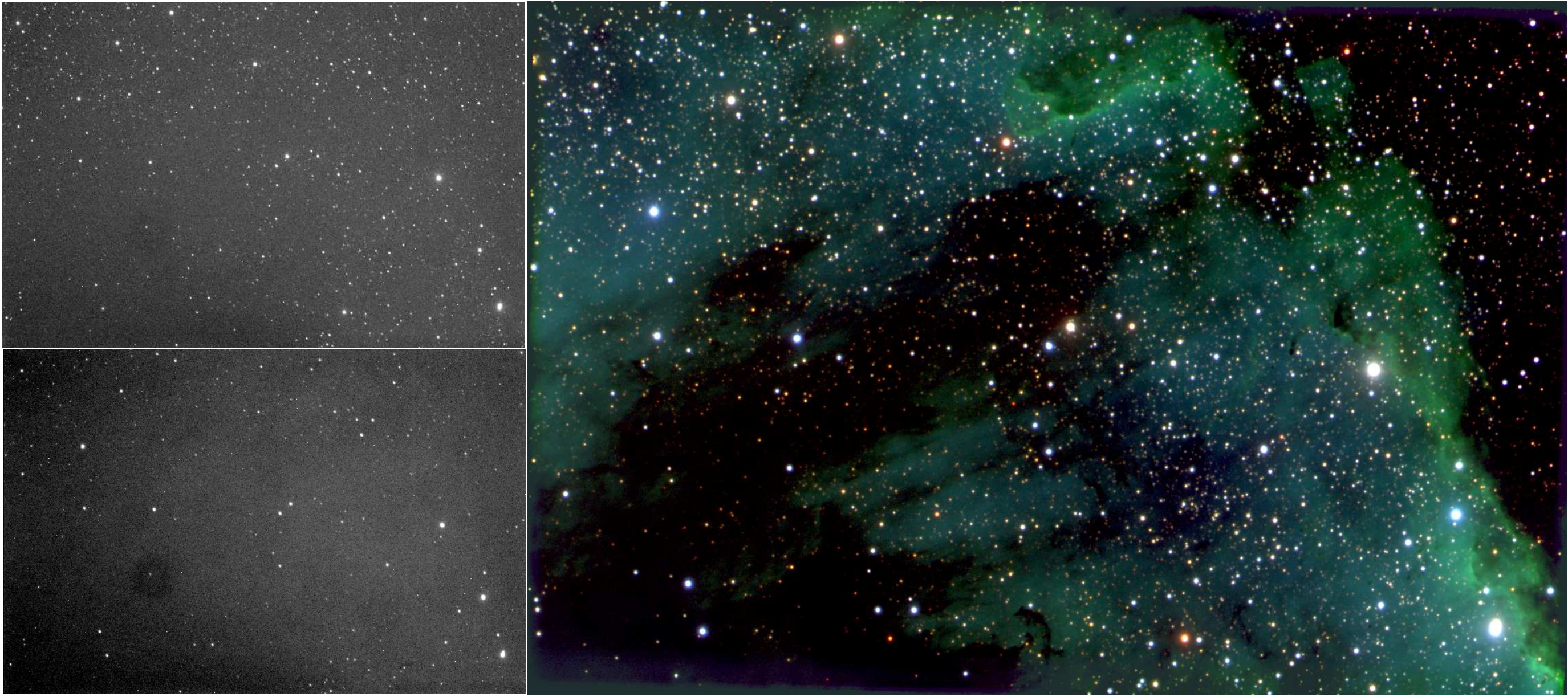}
\end{center}
\caption 
{ \label{fig:example_reduction}
An example of reduced frames from BANZAI. The left shows some example raw frames and the right panel shows a stack of BANZAI processed images.} 
\end{figure}

\begin{figure}
\begin{center}
\includegraphics[width=\textwidth]{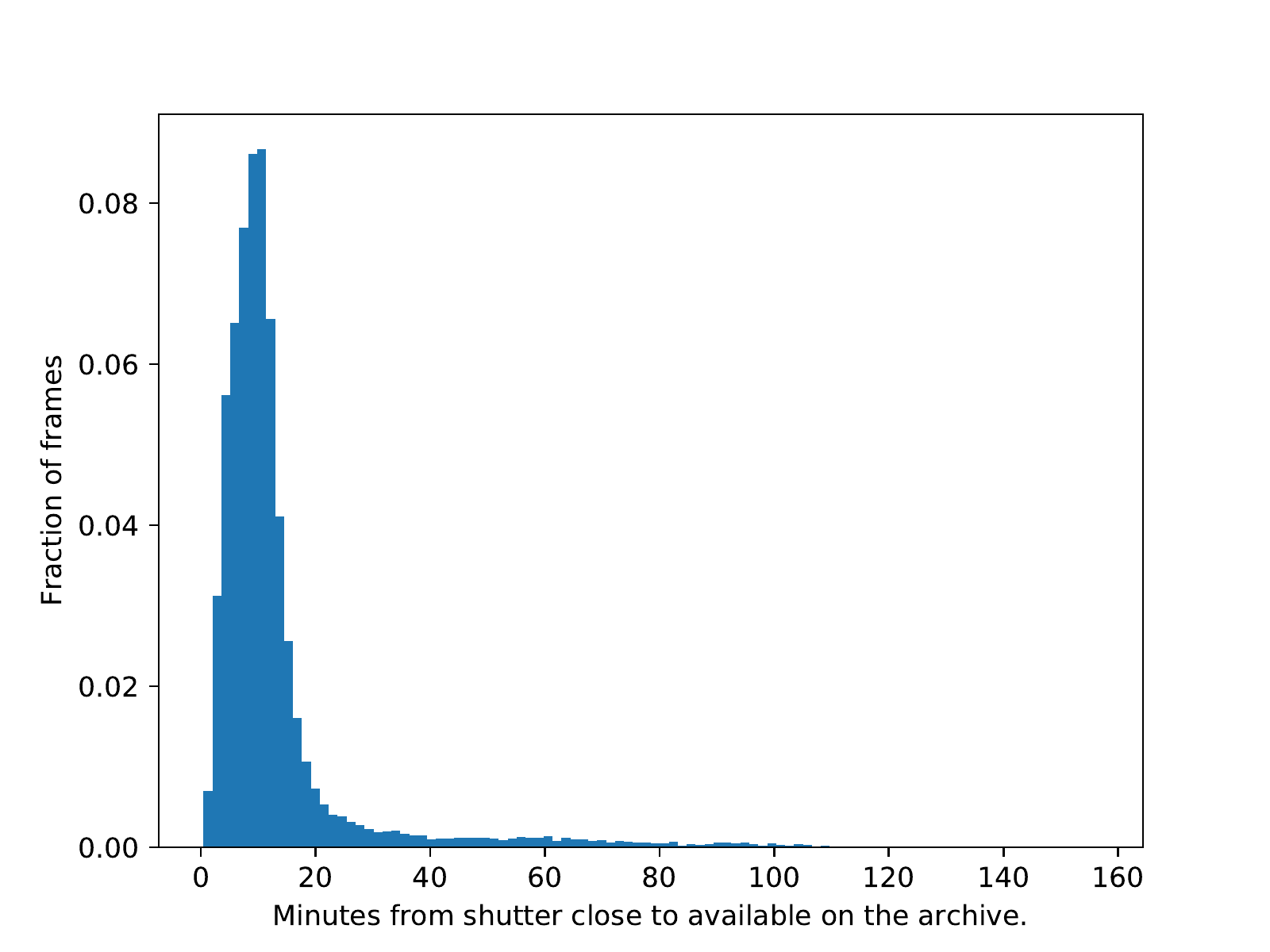}
\end{center}
\caption 
{ \label{fig:archive_lag}
Distribution of time between the end of the observation and availability on the archive. The median time between shutter close and being available to users is 10 minutes. Network issues and minor isolated pipeline failures drive the tail of the distribution to longer delays.} 
\end{figure}

BANZAI does standard image processing on every image taken with LCOGT: we mask bad pixels, subtract the overscan, subtract a master bias frame and a master dark, divide out a master flat, extract photometry for the sources in the field, and solve for the astrometry. We use SEP\cite{sep} to do source extraction and Astrometry.net\cite{Lang10} to solve for the astrometry. Raw files compared to a stack of reduced images from BANZAI are shown in Figure \ref{fig:example_reduction}. The median time from shutter close to the ``preview`` data being available for download by users is 10 minutes (the full distribution is shown in Figure \ref{fig:archive_lag}). The tail to longer delays is often due to network issues. 

\begin{figure}
\begin{center}
\includegraphics[width=\textwidth]{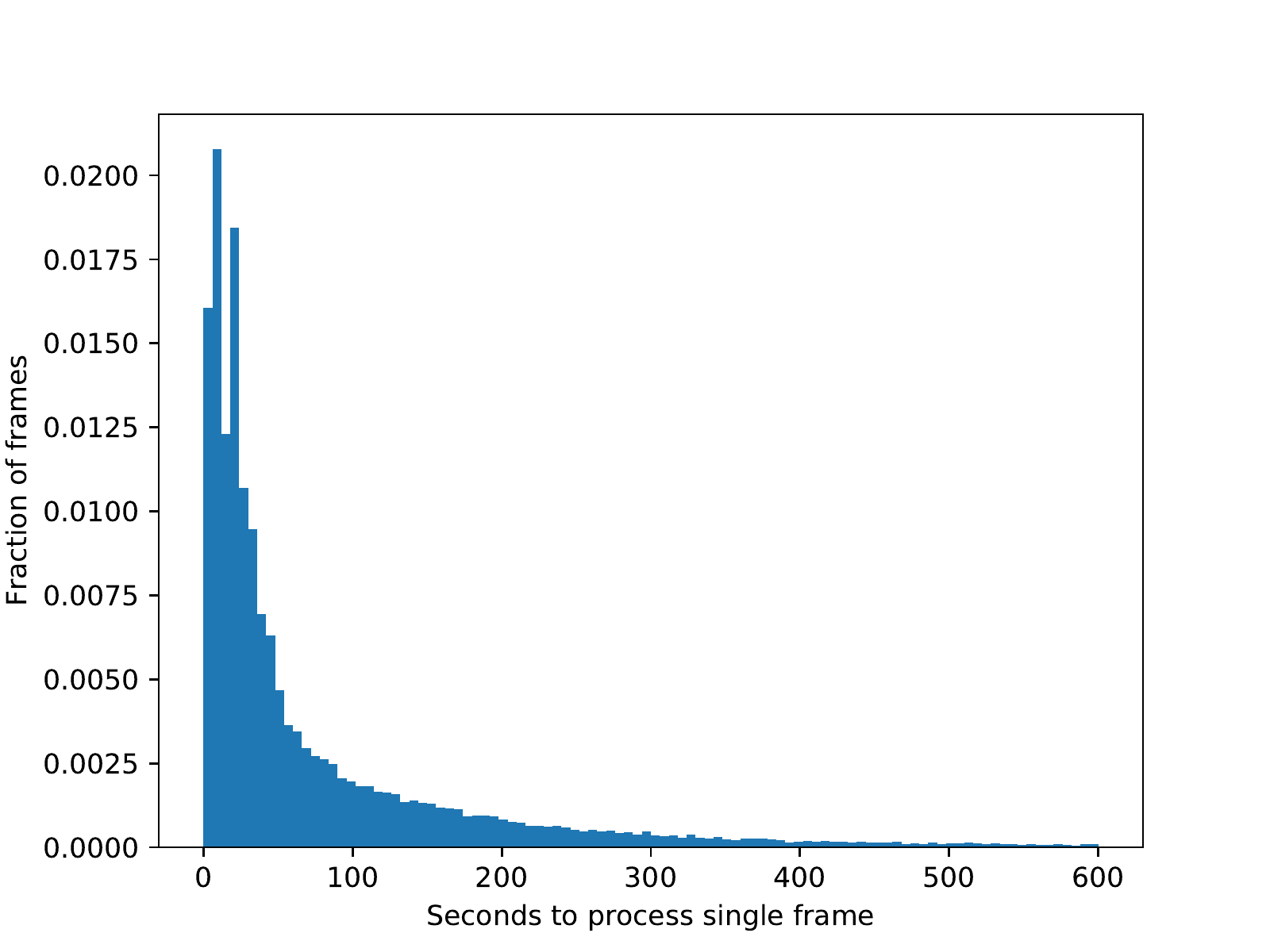}
\end{center}
\caption 
{ \label{fig:processing_lag}
Distribution of time it takes BANZAI to process an image. The median reduction time is 35~s. Dense fields like those taken near the galactic center for microlensing take longer to extract photometry and solve for the astrometry, driving the tail of the distribution to longer processing times.} 
\end{figure}

The median time for BANZAI to process an image is 2 minutes; Figure \ref{fig:processing_lag} shows the distribution of processing times for images. The long tail of processing times is due to dense fields, e.g. microlensing fields near the galactic center, that take longer to extract photometry and solve the astrometry than sparse fields.  

At the end of the local night for each site, we reprocess the images using master calibration images produced from calibrations taken during evening and morning twilight of that night. These constitute the science-quality reductions for users. By having both a preview and an end-of-night mode we satisfy users that need data quickly for rapidly evolving transients, but also use the best calibration data for the final reduction. 

\section{BANZAI architecture}
There are several architectural decisions that we made when developing BANZAI to minimize the barrier to entry for the code, to encourage transparency, and to allow for outside contributions from the broader astronomical community. The first choice is that we implemented BANZAI primarily in Python. Python has become one of the standard languages for astronomy, making it possible for a broad set of people to contribute to the development of the code base. The sections of the code that require high performance are written in Cython, with a few sorting algorithms implemented in C.

\subsection{Code Availability}
LCOGT is committed to being an open-source organization so the BANZAI pipeline is publicly available on GitHub. Being open-source encourages best practices like transparency, code readability, and maintainability. The open source community has also built a variety of useful tools like Coveralls to measure test coverage and ReadTheDocs\cite{readthedocs} that enables clear and easily accessible documentation. Because BANZAI is an open-source project, people outside of LCOGT can improve the quality of data that it produces because the code is not a proprietary black box. This improves the reproducibility of scientific results and allows for customizations for specific projects. By hosting BANZAI on GitHub, every commit is recorded, making every individual person's contributions apparent.

\subsection{Object model}
BANZAI is built as individual stages which are chained together to process images. The stage object is implemented as an abstract class which is used as the template for the Template Method pattern\cite{design_patterns}; the superclass contains all of the logic to pass images to the stage and for the top-level script to run the stage. The only piece that an individual stage needs to implement is a single method that takes a list of images and returns a list of processed images. The Template Method pattern employed here keeps the infrastructure for the pipeline simple and self-contained; the individual stages do not need to reimplement wrapper code, a common problem with pipelines. All of the images are kept in memory rather than being written to disk for each stage which keeps the file size small and the performance high. 

\subsection{Master calibrations}
Each night, every telescope in the LCOGT network takes calibration frames, e.g. biases, darks, and sky flats. BANZAI includes two utility template stage classes for these observations: a calibration stacker and a stage to correct science frames using a calibration stack. All of the file structure and retrieval logic for the most recent master calibration frames are built into the template classes. The specific stages, e.g., subtracting a dark frame, only have to implement the specialized logic to subtract the scaled dark from an image. Again the generic infrastructure is in a single location, making extension simple and limiting bugs from propagating via copy and paste. The master calibration metadata, such as the file path, are stored in a database. We use MySQL for our production instance but BANZAI itself is agnostic to the choice of database technology. We accomplish this by writing all of the queries in BANZAI using the Object Relational Model in SQLAlchemy\cite{sqlalchemy}. SQLAlchemy provides a unified frontend for most common database backends. This enables faster deployment and simpler testing; e.g. we use SQLite for testing and MySQL for production.

Tests on the master calibration frames have been crucial for high-quality results. If a bad calibration frame slips into the reductions (e.g.~a camera warms unexpectedly), images from multiple nights can be affected. To safeguard against this, we do two sets of comparisons on every calibration frame. The first is to compare the new calibration frames to previous ``good'' master calibration frames. A new ``good'' frame is created using the frames that pass this check. This allows the master calibration frames to evolve slowly over time, but rejects outliers. Prompt notification (early enough to allow intervention when needed) is an added benefit of real-time alerting. The second set of comparisons are done between all of the frames taken in the evening and morning twilight of an observing night. We employ a pixel-by-pixel comparison and reject any frames that fall outside the expected distribution (which is estimated using the median absolute deviation).

\subsection{Deployment}
The BANZAI code is deployed inside a Docker\cite{docker} container which allows rapid deployment on any machine as all of the dependencies are encapsulated in the container. If a user wants or needs to install the code locally instead of using Docker, the \textit{Dockerfile} acts as a recipe to install the necessary dependencies. Using Docker also allows us to run multiple instances of the pipeline in parallel, e.g. a preview mode that produces fast-turnaround reductions and an end-of-night mode that applies master calibration frames taken in the morning after a given observation.

The lifecycles of the Docker containers are managed by Rancher\cite{rancher}, a container orchestration platform. Rancher schedules CPU resources, creates and destroys containers, and maps network ports between containers. The frontend streamlines deploying upgrades: simply changing a version in a web form and clicking ``upgrade'' is all that is required to deploy a new version of the pipeline. Rancher also provides access to a terminal inside containers which can be used to debug issues. 

\begin{figure}
\begin{center}
\includegraphics[width=\textwidth]{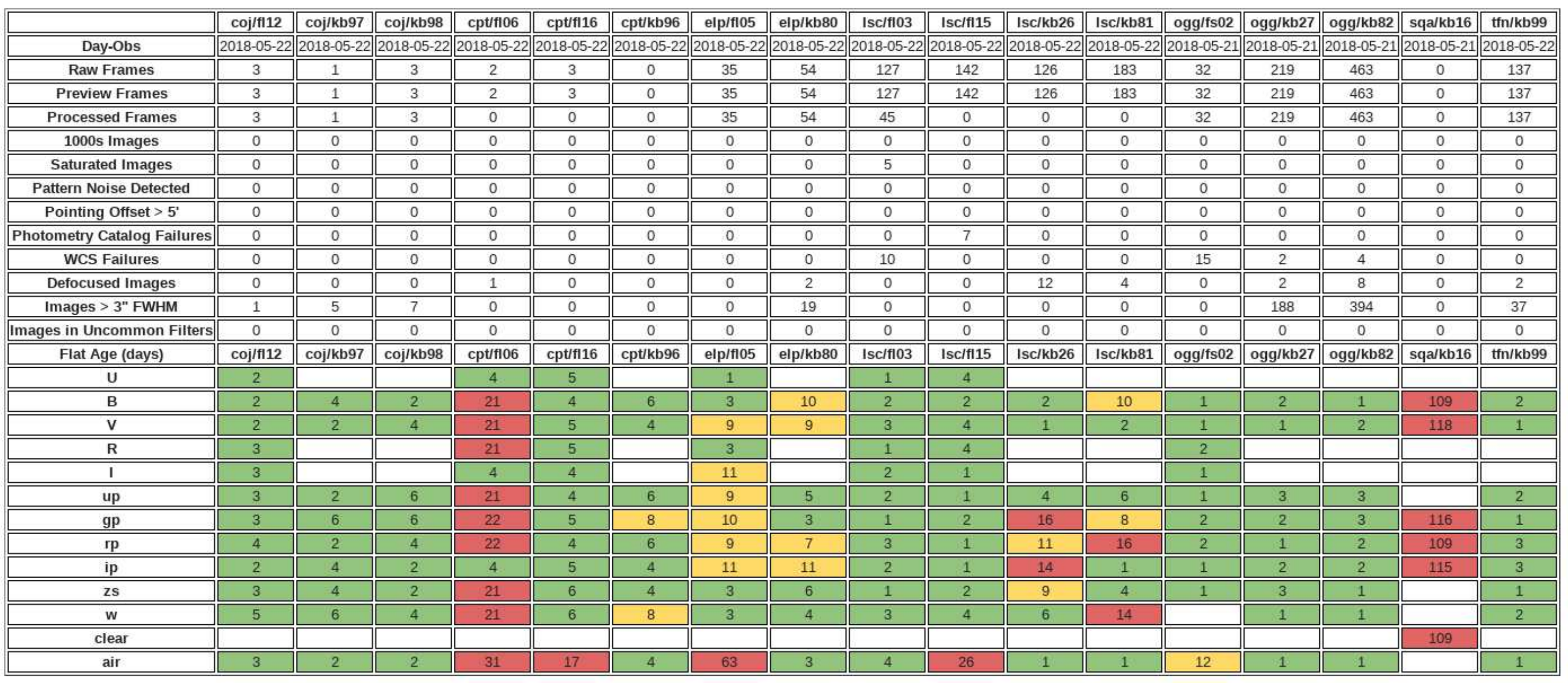}
\end{center}
\caption {\label{fig:dq_report} An example summary report on the state of data processing for the LCOGT network. Each column corresponds to an individual instrument. The top rows show the total number of images taken and processed from last night. The middle section shows the results from a variety of data quality tests. The bottom colored section shows the ages of flat-fields for common filters. Green denotes that the flat-fields are recent, less than a week old. Red shows filters for which a flat has not been taken for more than two weeks and require human intervention.} 
\end{figure} 

\subsection{Testing}
Unit tests for BANZAI are automatically run using continuous integration (CI) services: we use a public Travis-CI site\cite{travis} and an internal Jenkins\cite{jenkins} server. The Travis-CI page allows outside users to see the results of the unit tests, while the internal Jenkins server allows us to test deployment and to do integration and end-to-end tests that require large test datasets.

\section{Failure Detection and Performance Metrics}
One of the key challenges to run a data reduction service at this scale is detecting failures automatically. LCOGT produces thousands of images per night, more than is reasonable for any human being to check manually. Instead we have to rely on automated checks to detect and, if necessary, reject failed frames. The results of these checks and performance metrics are stored in ElasticSearch\cite{elasticsearch}, a NoSQL database. The NoSQL format for ElasticSearch makes it simple to add metrics or change the format of stored results. ElasticSearch also provides a web API to query the results of our tests. Using the APIs, we have developed several monitoring tools. One is a summary report that is emailed to several members of the LCOGT operations team. An example table is shown in Figure \ref{fig:dq_report}. This report is designed to quickly visualize the age of the master calibration frames (less than a week old is shown in green) and to summarize the occurrence of known failure modes across the LCOGT network. 
We also use the metrics to generate real-time alerts. These alerts can be sent via email or to the operations team in Slack\cite{slack}, a real-time communication platform. As these alerts are generated less than 10 minutes after shutter close, issues can be addressed before the whole observing night is affected and errors are more quickly detected rather than needing to wait for errors to be reported by users.

\begin{figure}
\begin{center}
\includegraphics[width=\textwidth]{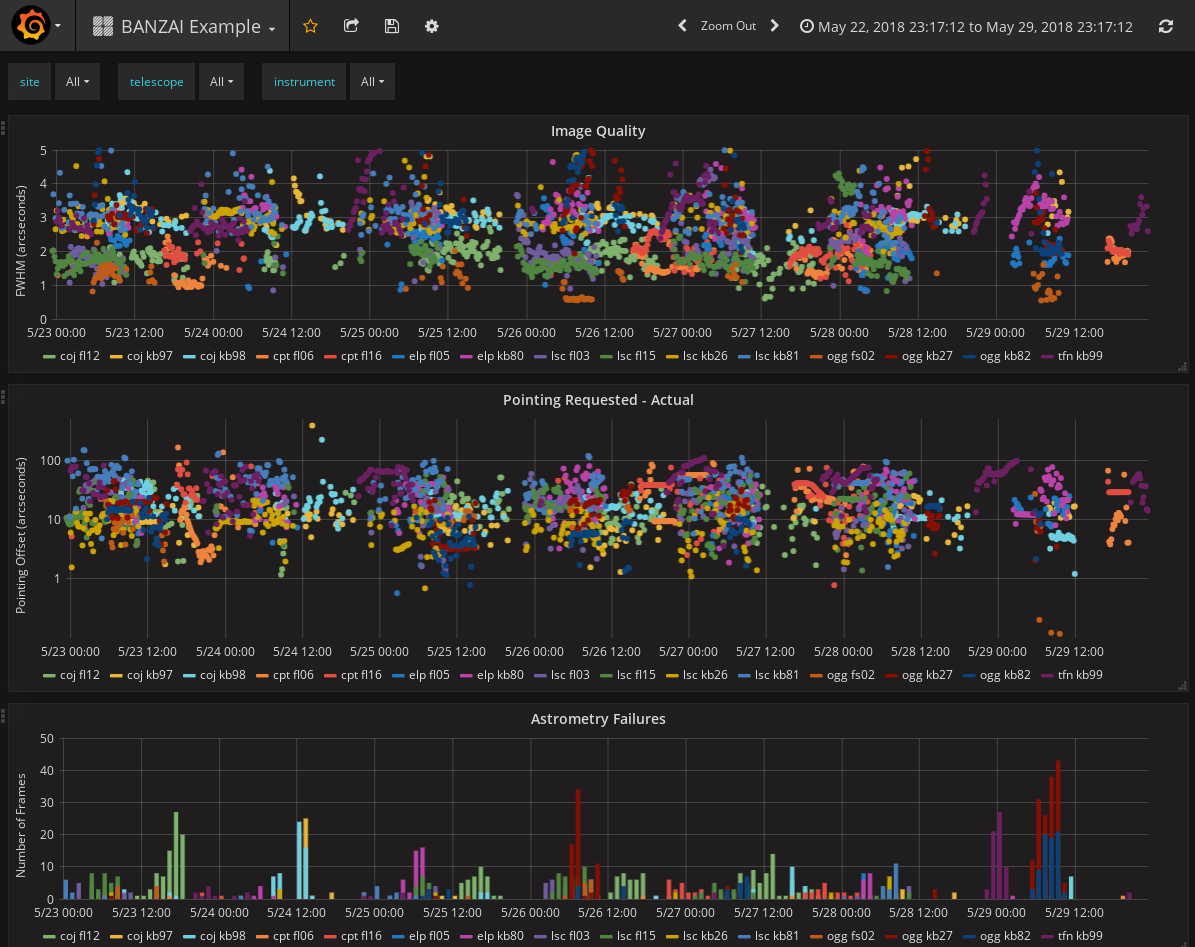}
\end{center}
\caption{\label{fig:grafana} An example dashboard that can be used to monitor the data quality being produced by the LCOGT network. The top panel shows the FWHM for each of our telescopes. Individual sites or cameras (shown with different colored points) can be selected using the dropdown menus at the top. The middle panel shows the offset between the requested center of the frame and the actual pointing. The bottom panel shows the number of frames per telescope that were not able to solve for the astrometry. The astrometry failure rate for BANZAI is 12.4\% for the last six months. About half of those failures are due to focus issues on the 0.4-m telescopes. The rest are typically due to the sky transparency being too low.} 
\end{figure} 

We also use metrics calculated by BANZAI to preform long-term monitoring of the network (e.g.~the optical performance of LCOGT\cite{Harbeck2018}). As with the summary reports, we can use the ElasticSearch APIs to visualize data-quality metrics. We use Grafana\cite{grafana} to plot time series of metrics based on image quality, telescope pointing, etc. Figure \ref{fig:grafana} shows an example dashboard with the Full Width Half Maximum (FWHM), the pointing quality, and the number of astrometric solution failures from the last week. A few example data-quality tests are described below.

There are two main failure modes of the cameras at LCOGT. The first is when the camera produces an image that is only electric noise around the value of 1000 (this value is an artifact of the firmware). Frames are automatically rejected if a significant fraction of the image exactly equals 1000. When this happens, an alert is posted to Slack so that the operations team can restart the camera.

The second failure mode comes from electrical interference. To detect this, we employ a 2D Fast Fourier Transform (FFT) and check for excess power due to pattern artifacts in the frame. This is recorded in ElasticSearch but does not currently generate an alert. Tuning the threshold to detect this pattern noise is still a work in progress to avoid false positives. This was producing an overwhelming number of alerts, making them counterproductive. \textit{Limiting alert fatigue is essential for monitoring a dataflow of this size.}

Detecting and recovering from failures is an area of active development.

\section{Conclusion}
Data reduction at the scale of the LCOGT network is not without challenges: reducing thousands of images per night requires a high-performance pipeline, which was the motivation for BANZAI. Our median time to process an image is 35 seconds (see Figure \ref{fig:processing_lag}), and the median time between the shutter closing and the processed image being available to the user on the science archive is less than 10 minutes as shown in Figure \ref{fig:archive_lag}. 

As LCOGT is producing data 24 hours a day, 7 days a week, robust deployment is essential: BANZAI is deployed in a Docker container that is managed by Rancher, making it easy to upgrade and redeploy the pipeline as necessary. As new instruments come online, BANZAI needs to be able to adapt, so the code is designed with the idea of extensibility in mind. Monitoring a dataflow of this size without human intervention is another significant challenge. To solve this, we have tightly integrated the BANZAI pipeline into telescope operations via queues and APIs. We also send status and metrics to an ElasticSearch database that we use to produce summary plots and metrics to monitor the health of the telescope network.

As future surveys like ZTF and LSST begin, going from follow-up observations to scientific results will require that facilities be able to reduce data with minimal human interaction, but still have insight into the datastream. The lessons we have learned at LCOGT will help to make that possible.


\acknowledgments 
We thank Martin Norbury, Todd Boroson, and Stefano Valenti for discussion about the design of BANZAI. CM was supported by supported by NSF grant AST-1313484.


\bibliographystyle{spiebib}   
\bibliography{main}   


\end{document}